\documentstyle[amsfonts,pra,multicol,aps]{revtex}

\begin{document}
\draft
\preprint{}
\title{Classical information capacity of superdense coding}
\author{Garry Bowen \cite{email}}
\address{Department of Physics,
Australian National University,
Canberra, A.C.T. 0200, Australia.
}
\date{\today}
\maketitle

\begin{abstract}
Classical communication through quantum channels may be enhanced by sharing entanglement.  Superdense coding allows the encoding, and transmission, of up to two classical bits of information in a single qubit.  In this paper, the maximum classical channel capacity for states that are not maximally entangled is derived.  Particular schemes are then shown to attain this capacity, firstly for pairs of qubits, and secondly for pairs of qutrits.
\end{abstract}

\begin{multicols}{2}


Quantum information exhibits many features which do not have analogues in classical information theory \cite{bennett00}.  For this reason, when a quantum channel is used for communication, there exist a number of different capacities for the different types of information transmitted through the channel \cite{barnum98,adami97,lloyd,bennett99}.

{\it Superdense coding} (referred to in this paper simply as dense coding), first proposed by Bennett and Wiesner \cite{bennett92}, is where the transmission of classical information through a quantum channel is enhanced by shared entanglement between sender and receiver.  The classical information capacity for a channel where sender and receiver share entanglement has been called the {\it entanglement-assisted classical capacity} $C_{E}$ \cite{bennett99}.  The classical capacity for dense coding, denoted here by $C$, provides a lower bound on $C_{E}$.

For Completely General Dense Coding (CGDC) \cite{bose00}, the sender, Alice, and receiver, Bob, share qubits in the state $\rho_{AB}$, Alice may encode a message using a set of unitary transformations $\{ U_{A}^{k} \}$, with {\it a priori} probabilities $\{ p_{k} \}$, on her qubit.  Alice then sends her qubit to Bob, who decodes the message by doing joint measurements on both qubits.

For pure states of pairs of $D$ state systems, where $\rho_{AB} = |\Psi_{AB} \rangle \langle \Psi_{AB} |$, the channel capacity has been derived by both Hausladen, {\it et. al.} \cite{hausladen96}, and, Barenco and Ekert \cite{barenco95}, and was shown to be, $C=\log D + S(\rho_{B})$.  Here $S$ is the von Neumann entropy $S(\rho)= -{\mathrm Tr} \rho \log \rho$, where the logarithm is base 2.

Bose, Plenio and Vedral \cite{bose00} have further proven that if Alice's alphabet of operators is restricted to the set of the identity and three Pauli matrices, $U_{A}^{k} \in \{ I, \sigma_{x}, \sigma_{y}, \sigma_{z} \}$, then the capacity for a pair of qubits is maximized by setting $p_{k} = 1/4$.  This scheme was labeled by the authors as Special Dense Coding (SDC).

In this paper, a bound on the channel capacity for dense coding is derived for arbitrary sets of unitary operators on pairs of qubits.  It is shown that the scheme of SDC attains that bound.  Further, the proof for the case of pairs of qutrits is outlined, utilizing the higher dimensional analogue of SDC.


Suppose Alice and Bob share pairs of qubits in the state $\rho_{AB}$, and Alice is restricted to using unitary operators and sending her message as a product state of letters, then the maximal amount of classical information that may be transferred is given by the Kholevo bound \cite{kholevo73},
\begin{eqnarray}
C &=& \max_{ \{ U_{A}^{k} , p_{k} \} } \Bigg[ S\left( \sum_{k} p_{k} (U_{A}^{k} \otimes I_{B}) \rho_{AB} (U_{A}^{k} \otimes I_{B})^{\dag} \right) \nonumber \\
&&- \sum_{k} p_{k} S\left( (U_{A}^{k} \otimes I_{B}) \rho_{AB} (U_{A}^{k} \otimes I_{B})^{\dag} \right) \Bigg] .
\label{eqn:capacity}
\end{eqnarray}
This bound has been shown to be asymptotically attainable by using product state block coding \cite{hausladen96,schumacher97}.

As the operators $U_{A}^{k} \otimes I_{B}$ are unitary, applying one of the operators to $\rho_{AB}$ will not change the eigenvalues.  Hence the entropy, which depends only on the eigenvalues, of each summand in the second term of Eq. (\ref{eqn:capacity}) remains unchanged, and the second term reduces to $S(\rho_{AB})$.
To maximize the capacity we must therefore maximize the first term,
\begin{equation}
S( \rho'_{AB} ) = S \left( \sum_{k} p_{k} (U_{A}^{k} \otimes I_{B}) \rho_{AB} (U_{A}^{k} \otimes I_{B})^{\dag} \right).
\label{eqn:firstterm}
\end{equation}

A general density matrix of a two qubit bipartite system may be expanded as,
\begin{equation}
\rho_{AB} = \sum_{ij} \lambda_{ij} \sigma_{A}^{i} \otimes \sigma_{B}^{j} ,
\end{equation}
where the $\sigma$'s consist of a scaled version of the set of Pauli matrices and the identity, that is,
\begin{eqnarray}
\sigma^{0} &=& \frac{1}{2} I_{2} = \frac{1}{2} \left( \begin{array}{rr} 1 & 0 \\ 0 & 1 \end{array} \right) \\
\sigma^{1} &=& \frac{1}{2} \sigma_{x} = \frac{1}{2} \left( \begin{array}{rr} 0 & 1 \\ 1 & 0 \end{array} \right) \\
\sigma^{2} &=& \frac{1}{2} \sigma_{y} = \frac{1}{2} \left( \begin{array}{rr} 0 & -i \\ i & 0 \end{array} \right) \\
\sigma^{3} &=& \frac{1}{2} \sigma_{z} = \frac{1}{2} \left( \begin{array}{rr} 1 & 0 \\ 0 & -1 \end{array} \right) .
\end{eqnarray}
By linearity we can obtain the reduced density matrices of $\rho_{AB}$ and $\rho'_{AB}$, by tracing over the expansions,
\begin{eqnarray}
\rho_{B} &=& {\mathrm Tr}_{A} \left[ \rho_{AB} \right] \\
&=& {\mathrm Tr}_{A} \left[ \sum_{ij} \lambda_{ij} \sigma_{A}^{i} \otimes \sigma_{B}^{j} \right] \\
&=& \sum_{ij} \lambda_{ij} {\mathrm Tr}_{A} \left[ \sigma_{A}^{i} \right] \sigma_{B}^{j} \\
&=& \sum_{j} \lambda_{0j} \sigma_{B}^{j} \label{eqn:rhoBexpn} ,
\end{eqnarray}
where the trace of each of the Pauli matrices is zero.  Also,
\begin{eqnarray}
\rho'_{B} &=& {\mathrm Tr}_{A} \left[ \sum_{k} p_{k} (U_{A}^{k} \otimes I_{B}) \rho_{AB} (U_{A}^{k} \otimes I_{B})^{\dag} \right] \\
&=& \sum_{ij} \lambda_{ij} \sum_{k} p_{k} {\mathrm Tr}_{A} \left[ U_{A}^{k} \sigma_{A}^{i}(U_{A}^{k})^{\dag} \right] \sigma_{B}^{j} \\
&=& \sum_{j} \lambda_{0j} \sigma_{B}^{j} \\
&=& \rho_{B} ,
\label{eqn:rhoprime}
\end{eqnarray}
using the fact that the trace of a matrix does not change under unitary transformations.

Combining the above derivations leads to the main result of this paper.  The amount of information that may be transferred for any $\{ U_{A}^{k}, p_{k} \}$ using an arbitrary, two qubit mixed state $\rho_{AB}$, is given by,
\begin{eqnarray}
C &=& S(\sum_{k} p_{k} (U_{A}^{k} \otimes I_{B}) \rho_{AB} (U_{A}^{k} \otimes I_{B})^{\dag}) \nonumber \\
&&- \sum_{k} p_{k} S((U_{A}^{k} \otimes I_{B}) \rho_{AB} (U_{A}^{k} \otimes I_{B})^{\dag}) \label{eqn:main1} \\
&=& S(\rho'_{AB} ) - S(\rho_{AB}) \label{eqn:main2} \\
&\leq& S(\rho'_{A}) + S(\rho'_{B}) - S(\rho_{AB}) \label{eqn:main3} \\
&\leq& \log 2 + S(\rho_{B}) - S(\rho_{AB}) . \label{eqn:main4}
\end{eqnarray}
Here, Eq. (\ref{eqn:main2}) follows from the discussion following Eq. (\ref{eqn:capacity}), and the first term rewritten as for Eq. (\ref{eqn:firstterm}).  Eq. (\ref{eqn:main3}) uses the subadditivity of the entropies of a bipartite system, and Eq. (\ref{eqn:main4}) from the relations $S(\rho'_{B}) = S(\rho_{B})$, by Eq. (\ref{eqn:rhoprime}), and the bound $S(\rho'_{A}) \leq \log 2$ for a qubit.

This bound is attainable using Special Dense Coding (SDC), where Alice uses the operators $U_{A}^{k} = 2\sigma_{A}^{k}$, each occurring with {\it a priori} probability $p_{k} = 1/4$.  Using this scheme, the state received by Bob is completely disentangled, that is,
\begin{eqnarray}
\rho'_{AB} &=& \sum_{k} p_{k} U_{A}^{k} \left( \sum_{ij} \lambda_{ij} \sigma_{A}^{i} \otimes \sigma_{B}^{j} \right) (U_{A}^{k})^{\dag} \label{eqn:SDCcap1} \\
&=& \sum_{ij} \lambda_{ij} \left( \sum_{k} \sigma_{A}^{k} \sigma_{A}^{i} \sigma_{A}^{k} \right) \otimes \sigma_{B}^{j} \label{eqn:SDCcap2} \\
&=& \sum_{j} \lambda_{0j} \sigma_{A}^{0} \otimes \sigma_{B}^{j} \label{eqn:SDCcap3} \\
&=& \frac{1}{2}I_{A} \otimes \rho_{B} \label{eqn:SDCcap4} ,
\end{eqnarray}
where Eq. (\ref{eqn:SDCcap3}) follows from Eq. (\ref{eqn:SDCcap2}) due to the relationship $\sigma^{j} \sigma^{i} \sigma^{j} = \frac{1}{2} \delta_{ij} \sigma^{j} - \frac{1}{4} \sigma^{i} = \pm \frac{1}{4} \sigma^{i}$, for $i,j \in \{ 1,2,3 \}$, and Eq. (\ref{eqn:SDCcap4}) is obtained by comparing Eq. (\ref{eqn:SDCcap3}) with Eq. (\ref{eqn:rhoBexpn}).  Thus, the capacity for SDC is equal to the bound given in Eqs. (\ref{eqn:main1})-(\ref{eqn:main4}), and SDC has been shown to be an optimal method for CGDC.

A similar result applies for two qutrits, where Alice uses the operators,
\begin{eqnarray}
U_{0} &=& \left( \begin{array}{ccc} 0 & 0 & 1 \\ 1 & 0 & 0 \\ 0 & 1 & 0 \end{array} \right) \\
U_{1} &=& \left( \begin{array}{ccc} 0 & 1 & 0 \\ 0 & 0 & 1 \\ 1 & 0 & 0 \end{array} \right) \\
U_{2} &=& \left( \begin{array}{ccc} 1 & 0 & 0 \\ 0 & e^{i\frac{2}{3}\pi} & 0 \\ 0 & 0 &  e^{i\frac{4}{3}\pi} \end{array} \right) \\
U_{3} &=& \left( \begin{array}{ccc} 1 & 0 & 0 \\ 0 & e^{i\frac{4}{3}\pi} & 0 \\ 0 & 0 &  e^{i\frac{2}{3}\pi} \end{array} \right) \\
U_{4} &=& -\frac{i}{\sqrt{3}} [ U_{0}, U_{2} ] \\
U_{5} &=& \frac{i}{\sqrt{3}} [ U_{0}, U_{3} ] \\
U_{6} &=& \frac{i}{\sqrt{3}} [ U_{1}, U_{2} ] \\
U_{7} &=& -\frac{i}{\sqrt{3}} [ U_{1}, U_{3} ] \\
U_{8} &=& I_{3} ,
\end{eqnarray}
with {\it a priori} probability $p_{j} = 1/9$, where $[U_{i},U_{j}]$ denotes the commutator.  Expanding the density matrix $\rho_{AB}$ in terms of the identity and the traceless Hermitian generators $\{ \lambda_{i} \}$ of $SU(3)$ \cite{caves00}, we find,
\begin{eqnarray}
    \rho'_{AB} &=& \sum_{j} p_{j} U_{j} \rho_{AB} U_{j}^{\dag} \\
    &=& \frac{1}{3}I_{3} \otimes \rho_{B} ,
\end{eqnarray}
and the capacity is given by,
\begin{equation}
C = \log 3 + S(\rho_{B}) - S(\rho_{AB}) .
\end{equation}

Similar constructions for arbitrary $N \times M$ state systems may easily be considered using analogues of the unitary transformations used in SDC.  The transformations consist of the set of cyclic permutations of the $D_{A}$ basis states of ${\mathcal H}_{A}$, where $D_{A}$ is the dimension of the Hilbert space ${\mathcal H}_{A}$ of Alice's state, the set of unitary matrices derived from the cyclic group generated by the matrix consisting of the $D_{A}$ roots of unity on the diagonal (up to overall phase), and the normalized commutators between elements of these two sets of transformations.  The connection between sets of unitary depolarizers, the existence of orthonormal bases of maximally entangled states, and dense coding has previously been noted by Werner \cite{werner00}.

We thus make the conjecture that for an $N \times M$ state system $\rho_{AB}$, the dense coding capacity is given by,
\begin{equation}
C = \log D_{A} + S(\rho_{B}) - S(\rho_{AB}) ,
\end{equation}
with $D_{A} = N$.


The result obtained in this paper agrees with the previously obtained results in the case of pure states.  The capacity may also be rewritten in the form,
\begin{eqnarray}
C(\rho_{AB}) &=& \log D_{A} - S(\rho_{A}) \nonumber \\
&&+ S(\rho_{A}) + S(\rho_{B}) - S(\rho_{AB}) \\
&=& C(\rho_{A}) + S(A,B) ,
\label{eqn:mutual}
\end{eqnarray}
for $C(\rho_{A})$ the capacity of sending qubit $A$ without access to qubit $B$, and $S(A,B)=S(\rho_{A})+S(\rho_{B})-S(\rho_{AB})$ the von Neumann mutual entropy of $\rho_{AB}$.  In this way it is shown that the capacity due to the joint measurement of both qubits is enhanced over the use of a single qubit by a factor equal to the von Neumann mutual entropy of the combined state.

The capacity also gives an exact bound on the mixedness of a state for when dense coding with that state may be said to fail \cite{vedral00}.  For an arbitrary bipartite state the capacity will not exceed $\log N$, for an $N \times M$ state system, with $N,M \in \{ 2,3 \}$, whenever $S(\rho_{B}) - S(\rho_{AB}) \leq 0$.  Disentangled states \cite{vedral98} satisfy the inequality,
\begin{equation}
S(\rho_{AB}) \geq \max \{ S(\rho_{A}) , S(\rho_{B}) \} ,
\label{eqn:disentangled}
\end{equation}
and therefore cannot be used to transmit more than $\log N$ bits per state.  The proof of Eq. (\ref{eqn:disentangled}) is given in Appendix \ref{sec:proof}.

It may also be noted \cite{private}, that the result for the capacity of a two qubit system also proves the conjecture that the capacity for dense coding is bounded by \cite{bose00},
\begin{equation}
C \leq 1 + E_D ,
\end{equation}
where $E_{D}$ is the (one way) distillable entanglement of $\rho_{AB}$, provided the Hashing Inequality \cite{horodecki00}, $E_{D} \geq S(\rho_{B}) - S(\rho_{AB})$, is true.

In summary, the classical information capacity of dense coding, through a noiseless channel, using arbitrary mixed states of two qubits or two qutrits has been derived.  A method of generalizing to $N \times M$ state systems has been outlined, and a conjecture made about the classical capacity of dense coding using such systems.


The author would like to thank Craig Savage and Tim Ralph for discussions.  The author is supported by the DETYA.

\appendix
\section{}
\label{sec:proof}

{\it Proof of} Eq. (\ref{eqn:disentangled}).  Suppose $\rho_{AB}$ is disentangled, then the density matrix may be written in the form $\rho_{AB} = \sum_{i} p_i \, \omega_{A}^{i} \otimes \omega_{B}^{i}$, with $\sum_{i} p_{i} = 1$ and $p_{i} > 0$, where the reduced density matrices $\omega^{i}$ are all pure states.  By the convexity of the expression $S(\rho_{B})-S(\rho_{AB})$ \cite{lieb73}, we have,
\begin{eqnarray}
S(\rho_{B}) - S(\rho_{AB}) &\leq& \sum_{i} p_i S(\omega_{B}^{i}) - \sum_{i} p_i S(\omega_{A}^{i} \otimes \omega_{B}^{i}) \nonumber \\
&=& 0 \nonumber
\end{eqnarray}
and hence $S(\rho_{AB}) \geq S(\rho_{B})$.  Similarly for $S(\rho_{AB}) \geq S(\rho_{A})$.

\end{multicols}

\end{document}